\documentclass{PoS}

\usepackage{grffile}

\def\simge{
    \mathrel{\rlap{\raise 0.511ex
        \hbox{$>$}}{\lower 0.511ex \hbox{$\sim$}}}}
\def\simle{
    \mathrel{\rlap{\raise 0.511ex 
        \hbox{$<$}}{\lower 0.511ex \hbox{$\sim$}}}}

\title{HISQ 2+1+1 light quark hadronic vacuum polarization at the physical point}

\ShortTitle{HISQ 2+1+1 light quark hadronic vacuum polarization at the physical point}

\author{\speaker{T. Blum}\\
        UConn\\
        E-mail: \email{thomas.blum@uconn.edu}}

\author{C. Aubin\\Fordham University}
\author{M. Golterman\\San Francisco State University}
\author{C. Jung\\Brookhaven National Laboratory}
\author{S. Peris\\Universitat Aut\`onoma de Barcelona}
\author{C. Tu\\UConn}

\abstract{We report on the computation of the light quark vacuum polarization with 2+1+1 flavors of HISQ fermions at the physical point and its contribution to the muon anomalous magnetic moment. Three ensembles, generated by the MILC collaboration, are used to take the continuum limit. We compare our result with recent ones in the literature.}

\FullConference{The 36th Annual International Symposium on Lattice Field Theory - LATTICE2018\\
		22-28 July, 2018\\
		Michigan State University, East Lansing, Michigan, USA.}

\begin{document}

\section{Introduction}

Fermilab experiment E989 is currently measuring the anomalous magnetic moment of the muon ($a_\mu=(g-2)/2$) with the goal of reducing the error on the BNL E821~\cite{Bennett:2006fi} result by a factor of four. Lattice calculations of the hadronic contributions, like the one reported here, are crucial to obtain and cross-check the Standard Model value to the same accuracy in order to discover new physics or lay to rest the longstanding discrepancy between theory and experiment.

Using lattice QCD and continuum, infinite-volume (perturbative) QED, one can calculate the hadronic vacuum polarization (HVP) contribution to the muon anomalous magnetic moment~\cite{Blum:2002ii},
{\color{black}
\begin{equation}
\label{eq:1}
a_\mu^{\rm HVP} = \left(\frac{\alpha}{\pi}\right)^{2}\int_{0}^{\infty} 
d q^2 \, f(q^2)\,{\hat\Pi}(q^2).
\end{equation}
}
$f(q^{2})$ is known, and $\hat\Pi(q^{2})$ is the subtracted HVP, $\hat\Pi(q^{2})=\Pi(q^{2})-\Pi(0)$, computed directly on a Euclidean space-time lattice from the Fourier transform of the vector current two-point function,
\begin{eqnarray}
\Pi^{\mu\nu}(q) &=& \int d^4x\, e^{i q x}\langle j^\mu(x)j^\nu(0)\rangle~~~~~~{j^\mu(x)=\sum_{i} Q_i\bar\psi(x)\gamma^\mu\psi(x)}\\
&=&\Pi(q^2)(q^\mu q^\nu-q^2\delta^{\mu\nu}).
\end{eqnarray}
The form in the second equation is dictated by Lorentz and gauge symmetries. 

In the following it is convenient to use the time-momentum representation~\cite{Bernecker:2011gh} which results from interchanging the order of the Fourier transform and momentum integrals.
\begin{eqnarray}
\Pi(q^2) -\Pi(0) &=& \sum_t \left(
\frac{\cos{qt} -1}{q^2} +\frac{1}{2} t^2
\right) C(t),\\
C(t) &=& \frac{1}{3}\sum_{\vec x,i}\langle j_i(\vec x, t)j_i(0)\rangle,\\
 w(t) &=& 2 \alpha^2 \int_{0}^{\infty} \frac{d \omega}{\omega} f(\omega^2)\left[\frac{\cos{\omega t} -1}{(2\sin{(\omega/2)})^2} +\frac{t^2}{2}\right],
 \end{eqnarray}
 where $C(t)$ is the Euclidean time correlation function and (\ref{eq:1}) becomes
 \begin{eqnarray}
 \label{eq:t-m amu}
a_{\mu}^{\rm HVP}(T) &=& \sum_{t}^T  w(t) C(t),
\end{eqnarray}
where $a_{\mu}^{\rm HVP}$ is obtained in the limit $T\to \infty$.
Note the double subtraction~\cite{Bernecker:2011gh,Lehner:2015bga,Aubin:2015rzx} in the cosine term above:  1/2 $t^2$ cancels $\Pi(0)$ configuration-by-configuration, and the leading finite volume correction is killed by the ``-1". 

The calculation rests heavily on the use of noise reduction techniques developed by the RBC and UKQCD collaborations, including all-mode (AM) and full volume low-mode (LM) averaging (see~\cite{Blum:2012uh,Blum:2015you} and references therein). We take a moment to describe the low-mode structure of the staggered fermion Dirac operator which plays a central role. The staggered operator is the sum of hermitian and anti-hermitian parts, so it satisfies (even-odd ordering)
\begin{eqnarray}
M\left(
\begin{array}{c}
n_o \\ n_e
\end{array}
\right) 
=\left(
\begin{array}{cc}
 m  & M_{oe}\\
 M_{eo} &  m
\end{array}
\right)
\left(
\begin{array}{c}
n_o \\ n_e
\end{array}
\right) &=&
(m + i\lambda_n) \left(
\begin{array}{c}
n_o \\ n_e
\end{array}
\right), 
\end{eqnarray}
and similarly for the preconditioned operator $M^\dagger M$,
\begin{eqnarray}
\left(
\begin{array}{cc}
 m  & -M_{oe}\\
 -M_{eo} &  m
\end{array}
\right)
\left(
\begin{array}{cc}
 m  & M_{oe}\\
 M_{eo} &  m
\end{array}
\right)
\left(
\begin{array}{c}
n_o \\ n_e
\end{array}
\right)
&=& \\
\left(
\begin{array}{cc}
m^2  - M_{oe} M_{eo} & 0\\
0 & m^2  - M_{eo} M_{oe}
\end{array}
\right)
\left(
\begin{array}{c}
n_o \\ n_e
\end{array}
\right) &=&
(m^2+\lambda_n^2)
\left(
\begin{array}{c}
n_o \\  n_e
\end{array}
\right).
\end{eqnarray}
Eigenvectors of the preconditioned operator are eigenvectors of $M$ with squared magnitude eigenvalue,  and the even part can be obtained from the odd part, $n_e =\frac{ -i}{\lambda_n} M_{eo} n_o.$ The 
eigenvalues come in $\pm $ pairs: If $n=(n_o, n_e)$ is an eigenvector with eigenvalue $\lambda_n$, then $n_-=(-1)^x \psi_n (x) = (-n_o,n_e)$ is also an eigenvector with eigenvalue $-\lambda_n$.
\begin{eqnarray}
\left(
\begin{array}{cc}
m  & M_{oe}\\
 M_{eo} &  m
\end{array}
\right)
\left(
\begin{array}{c}
-n_o \\ n_e
\end{array}
\right) &=&
(m - i\lambda_n) \left(
\begin{array}{c}
-n_o \\ n_e
\end{array}
\right).
\label{eq:minus lambda evec}
\end{eqnarray}
Thus we can construct pairs of eigenvectors corresponding to $\pm i \lambda$ for each $\lambda^2$, $n_o$ in the full volume low-mode average.

The full-volume LMA takes advantage of the spectral decomposition of quark propagator that requires only two independent volume sums instead of a volume-squared sum in the correlation function. We employ a conserved current (minus the three-hop Naik term) which makes the ``meson fields" a bit more complicated, 
\begin{eqnarray}
J^\mu(x) &=&  -\frac{1}{2} \eta_\mu(x) \left(
\bar{\chi}(x+\hat{\mu}) U^\dagger_\mu(x)\chi(x)
~+~\bar{\chi}(x)U_\mu(x)\chi(x+\hat{\mu})\right)
\end{eqnarray}
and spectral decomposition of propagator,
\begin{eqnarray}
M^{-1}_{x,y} &=& \sum_n^{N_{\rm (low)}}
\frac{\langle x |n\rangle\langle n|y\rangle}{m+i\lambda_n}+
\frac{\langle x |n_-\rangle\langle n_-|y\rangle}{m-i\lambda_n}
\end{eqnarray}

\vskip -0.5cm
{\begin{eqnarray}
4 J_\mu(t_x)J_\nu(t_y)&=&\sum_{m,n}
\sum_{\vec x} \frac{\langle m|x+\mu\rangle U^\dagger_\mu(x)\langle x|n\rangle}{\lambda_m} 
\sum_{\vec y} \frac{\langle n|y\rangle U_\nu(y)\langle y+\nu|m\rangle}{\lambda_n} 
\\\nonumber
&+&
\sum_{\vec x} \frac{\langle m|x\rangle U_\mu(x)\langle x+\mu|n\rangle}{\lambda_m} 
\sum_{\vec y} \frac{\langle n|y\rangle U_\nu(y)\langle y+\nu|m\rangle}{\lambda_n} 
\\\nonumber
&+&
\sum_{\vec x} \frac{\langle m|x+\mu\rangle U^\dagger_\mu(x)\langle x|n\rangle}{\lambda_m} 
\sum_{\vec y} \frac{\langle n|y+\nu\rangle U_\nu^\dagger(y)\langle y|m\rangle}{\lambda_n} 
\\\nonumber
&+&
\sum_{\vec x} \frac{\langle m|x\rangle U_\mu(x)\langle x+\mu|n\rangle}{\lambda_m} 
\sum_{\vec y} \frac{\langle n|y+\nu\rangle U^\dagger_\nu(y)\langle y|m\rangle}{\lambda_n} ,
\end{eqnarray}
}
where $\lambda_n$ is shorthand for either $m\pm i\lambda_n$, and to compute the above we construct the meson fields
\begin{eqnarray}
(\Lambda_{\mu}(t))_{n,m} 
&=& \sum_{\vec x} {\langle n|x\rangle U_\mu(x)\langle x+\mu|m\rangle} (-1)^{(m+n)x+m},
\end{eqnarray}
(with eigenvector ordering $\lambda_0, -\lambda_0, \lambda_1, -\lambda_1, \dots, -\lambda_{2 N_{\rm low}}$).

\begin{figure}[htbp]
\begin{center}
\includegraphics[width=0.4\textwidth]{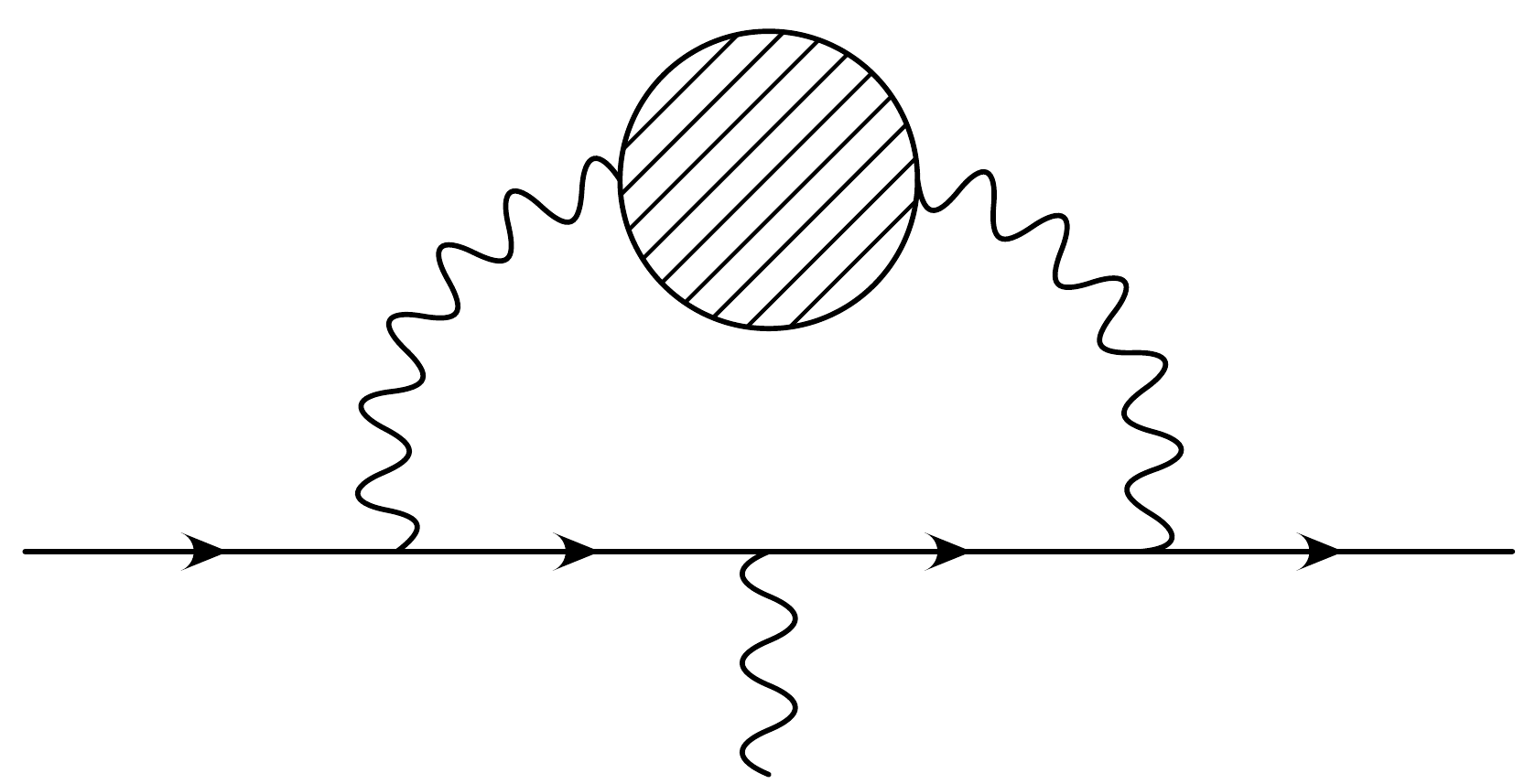}
\includegraphics[width=0.4\textwidth]{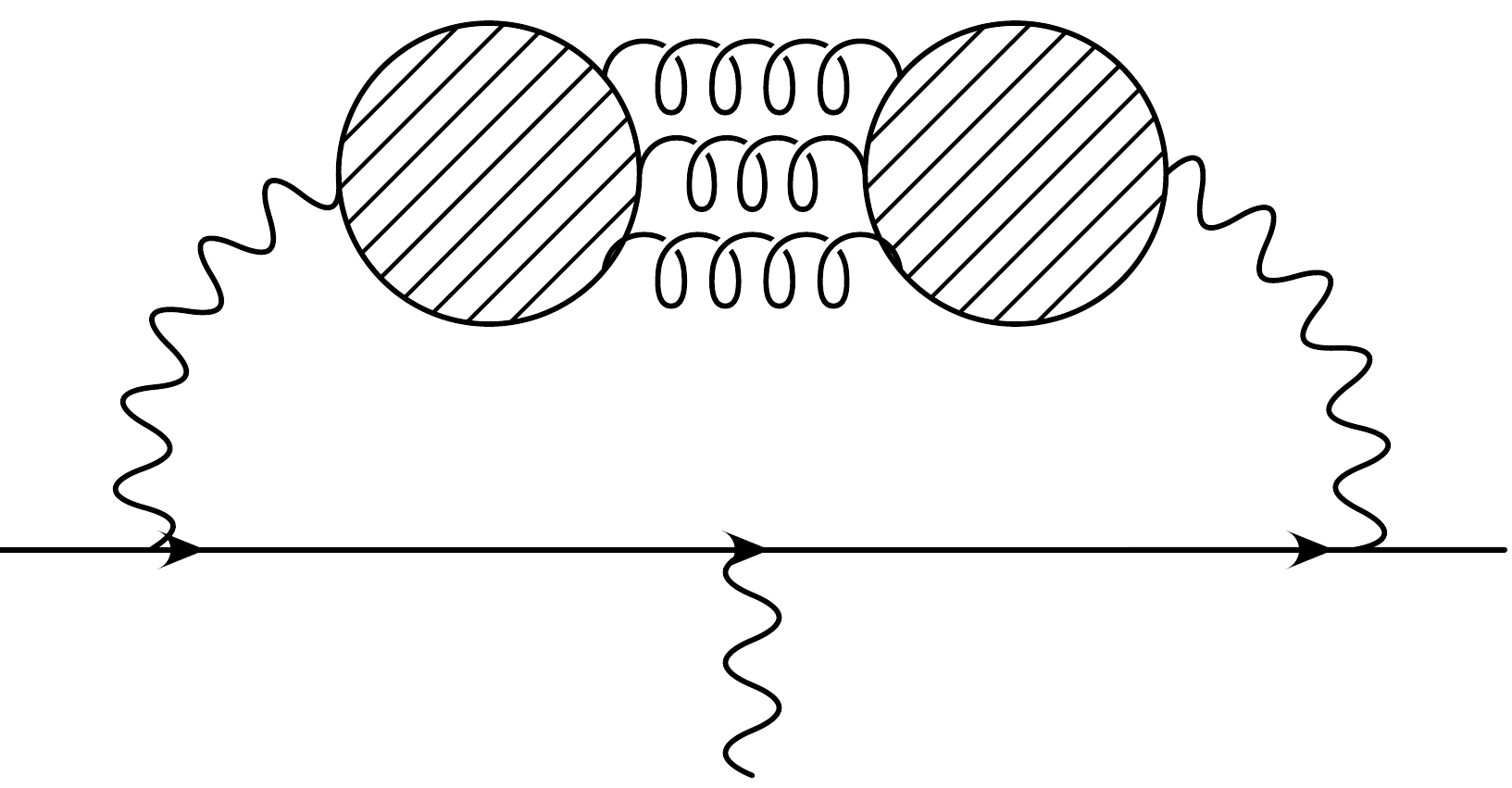}
\caption{The quark connected (left) and disconnected (right) diagrams contributing to the hadronic vacuum polarization contribution to the muon anomaly.}
\label{fig:vacpol}
\end{center}
\end{figure}

The total contribution to $a_\mu$ comes from both connected- and disconnected- quark line diagrams shown in Fig.~\ref{fig:vacpol}, for each flavor of quark in Nature. The u, d quark connected contributions are by far the largest, and we only compute them in this work. Comparison to other precise calculations will provide important validation for the lattice method.

\section{Preliminary Results}
We use the 2+1+1 flavor, physical mass ensembles generated by the MILC collaboration at three lattice spacings shown in Tab.~\ref{tab:ensembles}.
\begin{table}[htp]
\begin{center}
\begin{tabular}{|c|c|c|c|c|c|c|}
$m_\pi$ (MeV) & $a$ (fm) & size & $L$ (fm) & $m_\pi L$ & LM & meas (approx-exact-LMA) \\
\hline
133 & 0.12224(31) & $48^3\times 64$ & 5.87 & 3.9 & 3000 &26-26-26 \\
130 & 0.08786(26) &  $64^3\times 128$& 5.62 & 3.7 & 3000 &18-18-40 \\
135 & 0.05662(18) &  $96^3\times 192$& 5.44 & 3.7 & 2000 &14-22-18
\end{tabular}
\end{center}
\caption{Gauge field ensemble parameters\cite{PhysRevD.90.074509}. LM is the number of low-modes of the preconditioned Dirac operator. The number of configurations used for approximate, exact, and LMA measurements in this study are given in the last column.}
\label{tab:ensembles}
\end{table}%

\begin{figure}[htbp]
\begin{center}
\includegraphics[width=0.5\textwidth]{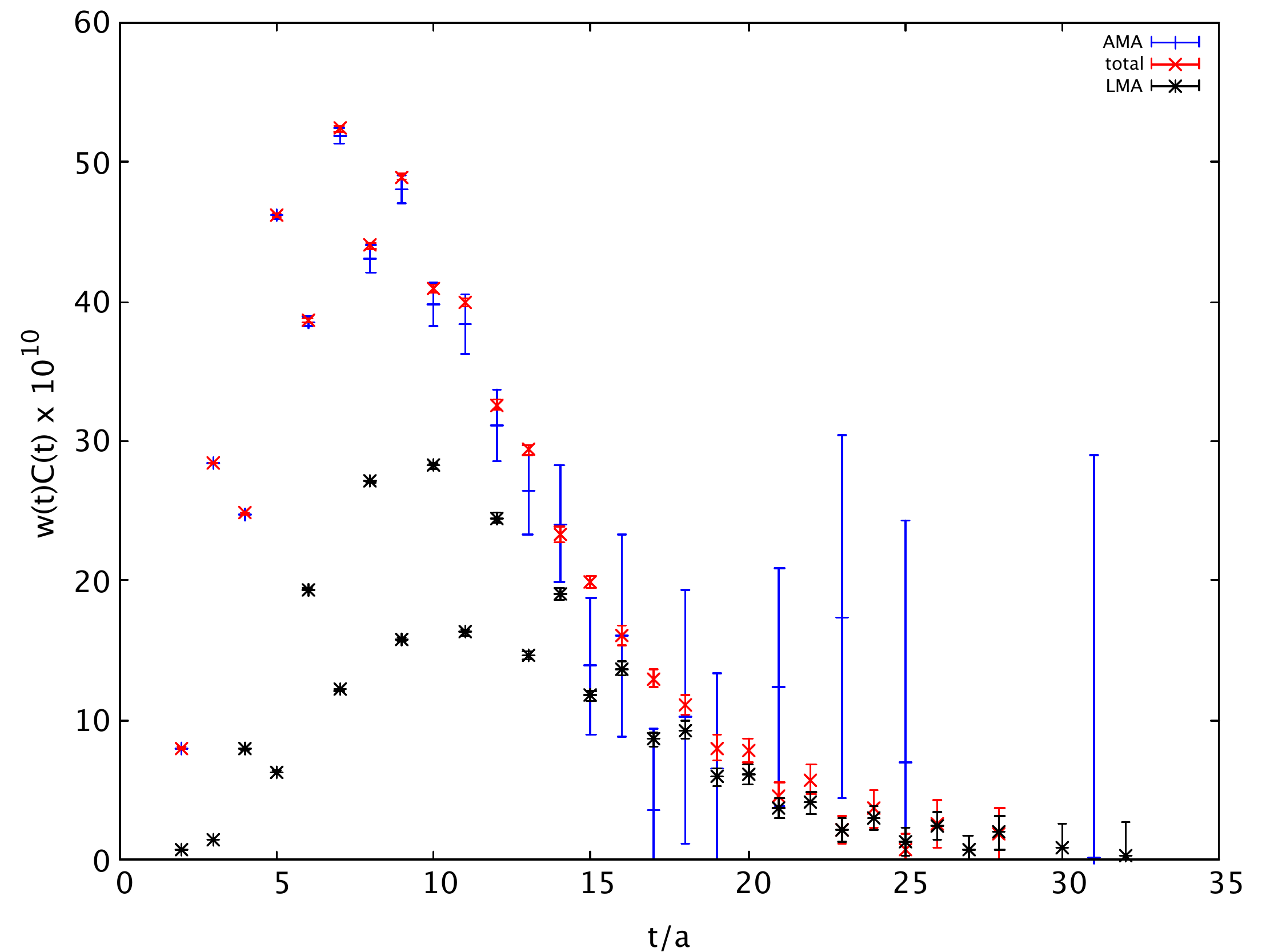}
\caption{The integrand (summand) in (\ref{eq:t-m amu}) for the first ensemble in Tab.~\ref{tab:ensembles}. Total (crosses), low-modes (stars), and AMA (plusses) contributions. Odd-parity, excited state oscillations are readily apparent.}
\label{fig:amu integrand}
\end{center}
\end{figure}

In Fig.~\ref{fig:amu integrand} the integrand in (\ref{eq:t-m amu}) computed on the $48^3$ ensemble is shown along with the full volume LMA (using $3000\times2$ low-modes per configuration) and AMA (256 sloppy and 8 exact point source propagators per configuration) contributions. 
\begin{eqnarray}
\label{eq:ama+lma}
\langle O\rangle &=& \langle O\rangle_{\rm exact} - \langle O\rangle_{\rm approx} + \frac{1}{N} \sum_{i} \langle O_i\rangle_{\rm approx}- \frac{1}{N} \sum_{i} \langle O_i\rangle_{\rm LM} 
+ \frac{1}{V} \sum_{i} \langle O_i\rangle_{\rm LM}.
\end{eqnarray}
As observed in~\cite{Blum:2018mom} there is a huge reduction in statistical error from the low-mode average, the last term in (\ref{eq:ama+lma}) (compare the total with and without LMA). The error reduction is especially large for large distance, as expected.

\begin{figure}[htbp]
\begin{center}
\includegraphics[width=0.475\textwidth]{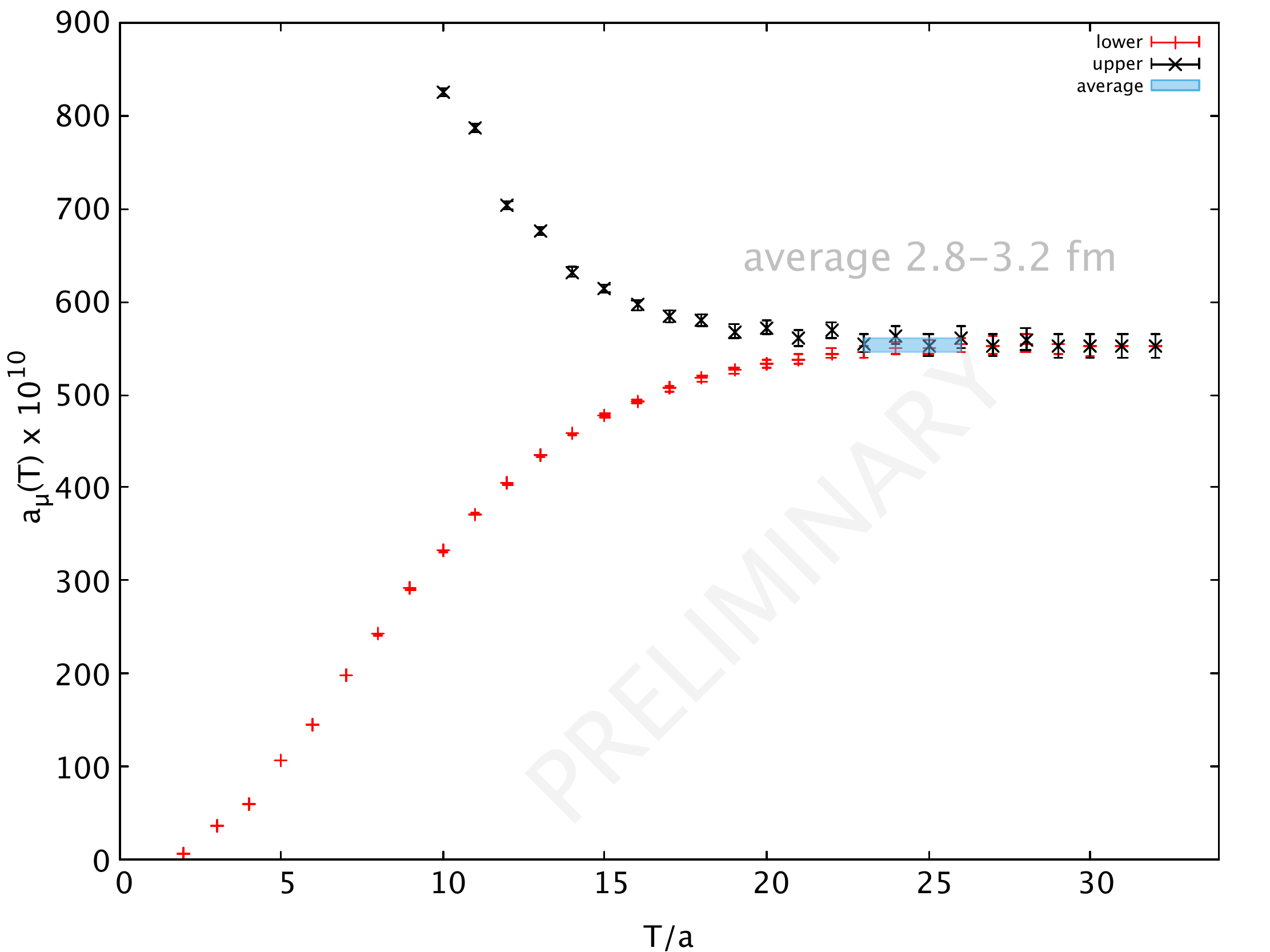}
\includegraphics[width=0.475\textwidth]{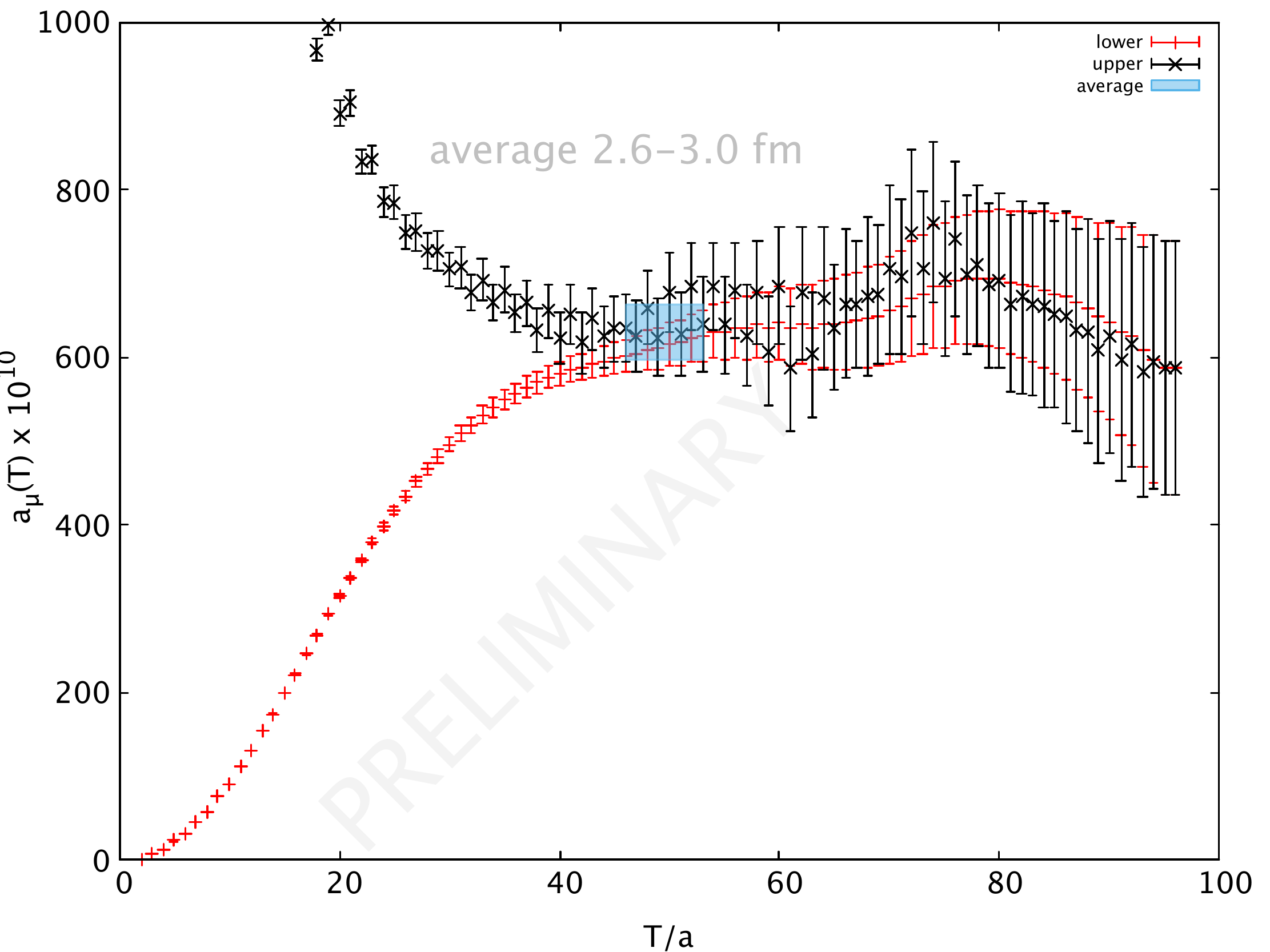}\\
\caption{Bounding method for total contribution to the muon anomaly. $48^3$ (left) and $96^3$ (right) ensembles.}
\label{fig:bound}
\end{center}
\end{figure}

In order to reduce further the statistical errors on the integrated result, we employ the bounding method~\cite{Blum:2018mom,Borsanyi:2017zdw} wherein $C(t)$, for $t>T$, is given by $C(t)=0$ (lower bound), and $C(t)=C(T) e^{-E_0(t-T)}$ (upper bound), where $E_0=2\sqrt{m_\pi^2+(2\pi/L)^2}$, $i.e$, the lowest energy state in the vector channel. At sufficiently large $T$ the bounds overlap, and an estimate for $a_\mu$ can be made which may be more precise than simply summing over the noisy long-distance tail. In Fig.~\ref{fig:bound} results are shown for the $48^3$ and $96^3$ ensembles. Central values for $a_\mu$ are averages over a suitable range where $T$ is large enough for the bounds to overlap but not so large that statistical errors blow up.  

\begin{figure}[htbp]
\begin{center}
\includegraphics[width=0.6\textwidth]{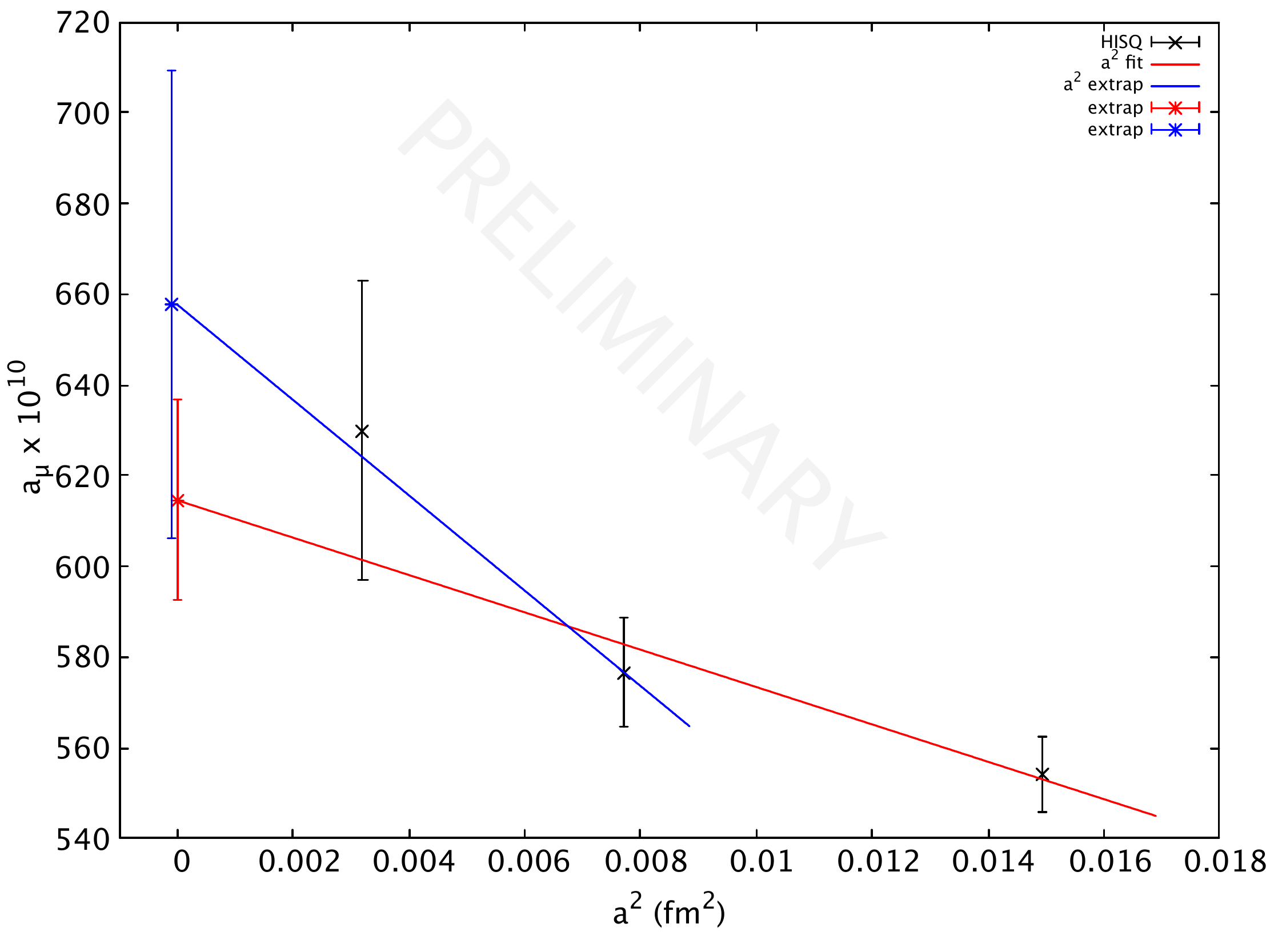}
\caption{Continuum limit of the muon anomaly at fixed volume from a linear fit and a linear extrapolation of the finest two points.}
\label{fig:cont lim}
\end{center}
\end{figure}

The muon anomaly for each lattice spacing is plotted in Fig.~\ref{fig:cont lim}. We do not gain much from the bounding method for the $48^3$ ensemble which has very small statistical errors already. But on the larger $96^3$ ensemble there is a clear advantage. The statistical errors in the latter case are larger likely because we have fewer measurements (see Tab.~\ref{tab:ensembles}) but also because fewer low-modes ($2\times2000$ compared to $2\times 3000$) were used on the $96^3$ lattice due to the expense.  We take a simple linear ansatz in $a^2$ , the leading artifact in the time-momentum representation, to take the continuum limit at fixed volume ($L\approx 5.5$ fm). The result is about $614\times 10^{-10}$ with a statistical error of $22\times10^{-10}$ which is consistent with recent results in the literature (see, $e.g.$, contributions to these proceedings), but with a relatively large error.

\begin{figure}[htbp]
\begin{center}
\includegraphics[width=0.475\textwidth]{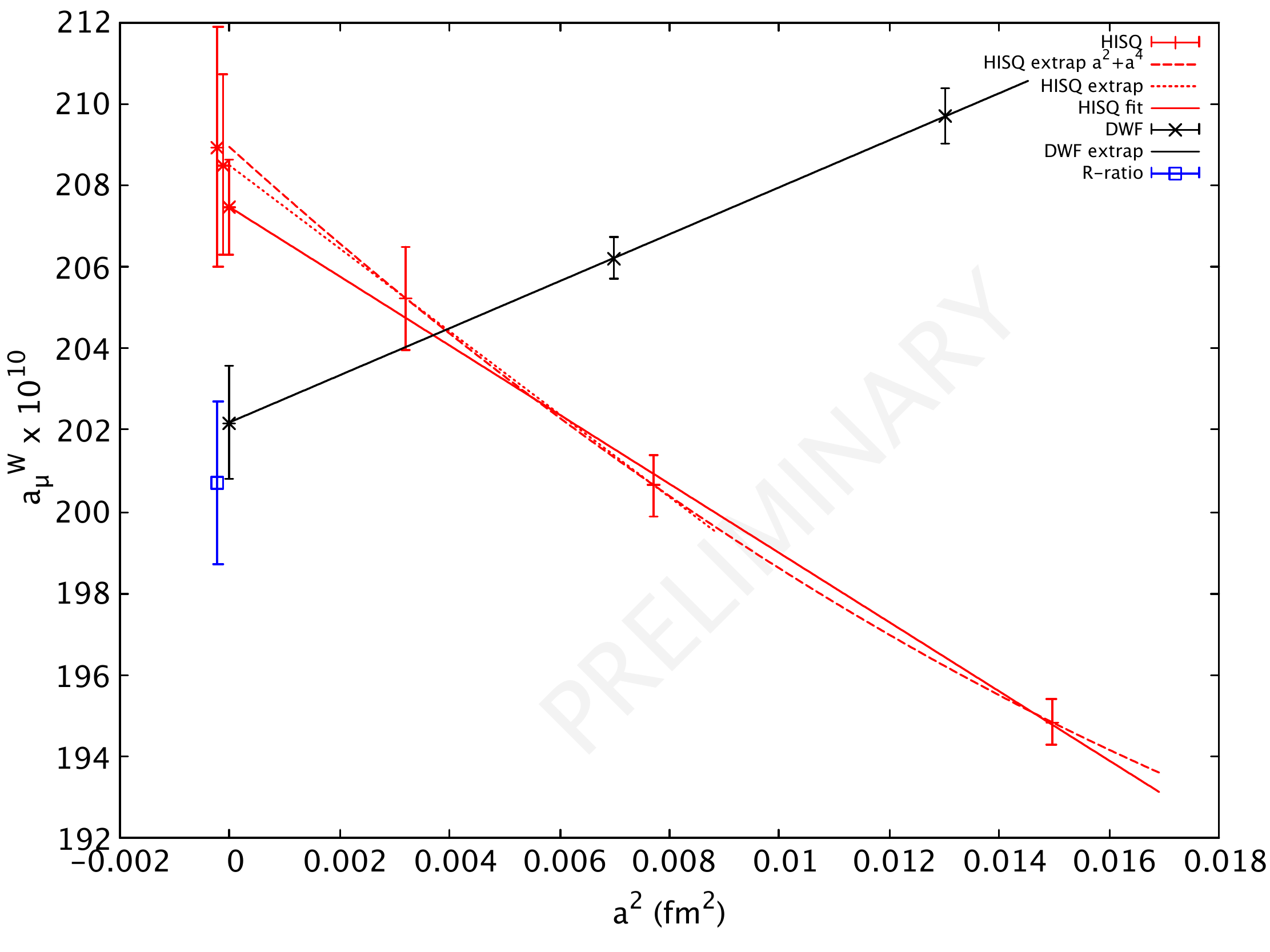}
\includegraphics[width=0.475\textwidth]{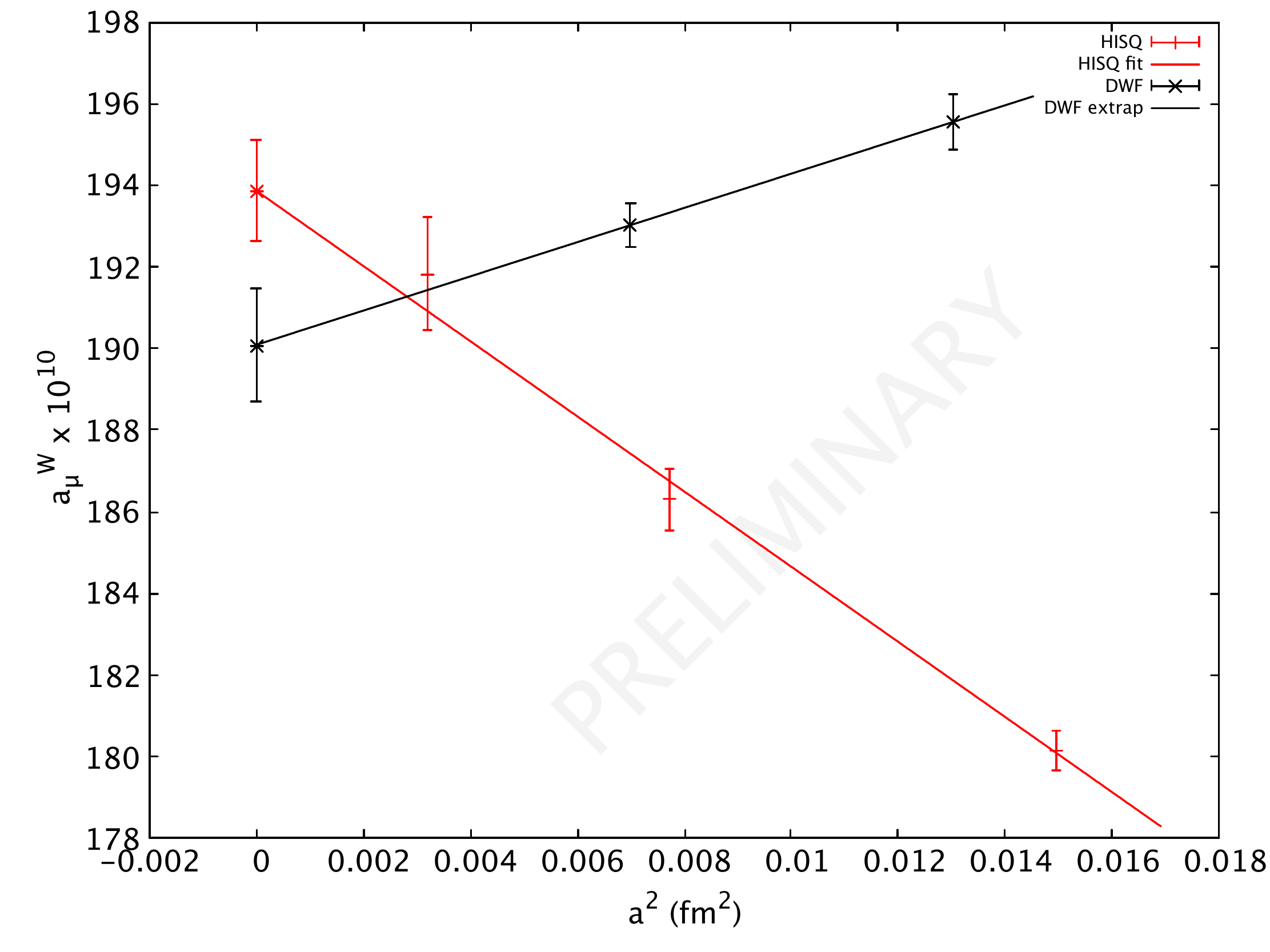}
\caption{Continuum limit combined with the window method. $t_0=0.4$ fm, $t_1=1$ fm, $\Delta=0.15$ (left) and 0.3 (right). DWF fermions~\cite{Blum:2018mom} (crosses) and HISQ fermions (plusses). In the left panel we show a linear fit (HISQ), linear extrapolations (both) and a quadratic extrapolation (HISQ).}
\label{fig:window}
\end{center}
\end{figure}

To explore a more precise comparison with other results, we adopt the window method of Ref.~\cite{Blum:2018mom}. 
\begin{equation}
a_\mu^{W} = \sum C(t) w(t) (\Theta(t,t_0,\Delta)-\Theta(t,t_1,\Delta)),~~~
\Theta(t,t',\Delta) = 0.5 (1+\tanh((t-t')/\Delta))
\end{equation}
where $t_1-t_0$ is the size of the window and $\Delta$ is a suitably chosen width that smears out the window at either edge. We choose windows to avoid both lattice artifacts at small distance and large statistical errors at large distance.
Figure~\ref{fig:window} displays results for two representative windows along with values from the recent RBC/UKQCD computation using domain wall fermions. The results should agree in the continuum limit up to small systematics. We also show the corresponding dispersive/$e^+e^-$ value. 

The HISQ result is 2-3 standard deviations above the DWF and dispersive ones. The largest difference is about $7\times 10^{-10}$, or roughly one percent of the total HVP contribution to $a_\mu$. Adding an $a^4$ term or leaving out the largest lattice spacing point tend to give even higher values with somewhat larger statistical errors. It is interesting to note that the HISQ and DWF lattice spacing errors are comparable. While we have not included estimates of the finite volume errors in our calculation, these are expected to be very small in the 0.4-1.0 fm window. Likewise, the absence of charm sea quarks in the DWF result is estimated from perturbation theory to be very small~\cite{Blum:2018mom}.

\section{Summary/Outlook}

We have presented a lattice QCD calculation of the light quark HVP contribution to the muon anomaly with 2+1+1 flavors of HISQ fermions. Three ensembles at the physical point were used to take the continuum limit at fixed volume ($L\approx 5.5$ fm), and the results are broadly consistent with those in the literature. Using the window method, a precise comparison yields a value that is bit higher than the dispersive result and a recent one using DWF. Statistics for the finer two ensembles are being improved and should illuminate any discrepancies or resolve them. Such comparisons are crucial for the upcoming confrontation with experiment E989 at Fermilab.

\section{Acknowledgments}
This work was partially supported by the US DOE. Computational resources were provided by the USQCD Collaboration. We thank the MILC collaboration for the use of their gauge configurations.

\bibliographystyle{JHEP}
\small
\bibliography{../talk/ref}

\end{document}